\documentclass[sigconf,dvipsnames]{acmart}
\usepackage{multirow}
\usepackage{colortbl}
\usepackage{makecell}
\usepackage[inline]{enumitem}

\newcommand{\set}[1]{\mathcal{#1}}

\newcommand{\popularitysearch}{\texttt{Pop$_{S}$}}
\newcommand{\popularityrecs}{\texttt{Pop$_{R}$}}
\newcommand{\recsmodel}{\texttt{Gen$_{R}$}}
\newcommand{\searchmodel}{\texttt{Gen$_{S}$}}
\newcommand{\jointmodel}{\texttt{Gen$_{R+S}$}}
\newcommand{\biencoder}{\texttt{Bi-encoder}}
\newcommand{\bm}{\texttt{BM25}}
\newcommand{\sone}{SIM1}
\newcommand{\stwo}{SIM2}
\newcommand{\sthree}{SIM3}

\theoremstyle{definition}
\newtheorem{hyp}{Hypothesis}[]

\acmConference[RecSys '24]{Proceedings of the 18th ACM Conference on Recommender Systems}{14–18 October 2024}{Bari, Italy}

\AtBeginDocument{%
  \providecommand\BibTeX{{%
    \normalfont B\kern-0.5em{\scshape i\kern-0.25em b}\kern-0.8em\TeX}}}

\copyrightyear{2024}
\acmYear{2024}
\setcopyright{acmlicensed}\acmConference[RecSys '24]{18th ACM Conference on Recommender Systems}{October 14--18, 2024}{Bari, Italy}
\acmBooktitle{18th ACM Conference on Recommender Systems (RecSys '24), October 14--18, 2024, Bari, Italy}
\acmDOI{10.1145/3640457.3688123}
\acmISBN{979-8-4007-0505-2/24/10}

\begin{document}

\title{Bridging Search and Recommendation in Generative Retrieval: Does One Task Help the Other?}


\author{Gustavo Penha$^1$, Ali Vardasbi$^1$, Enrico Palumbo$^2$, Marco de Nadai$^3$, Hugues Bouchard$^4$}
\affiliation{
\institution{Spotify}
\country{$^1$Netherlands, $^2$Italy, $^3$Denmark, $^4$Spain}
}
\email{{gustavop,aliv,enricop,mdenadai,hb}@spotify.com}  



\begin{abstract}

Generative retrieval for search and recommendation is a promising paradigm for retrieving items, offering an alternative to traditional methods that depend on external indexes and nearest-neighbor searches. Instead, generative models directly associate inputs with item IDs. Given the breakthroughs of Large Language Models (LLMs), these generative systems can play a crucial role in centralizing a variety of Information Retrieval (IR) tasks in a single model that performs tasks such as query understanding, retrieval, recommendation, explanation, re-ranking, and response generation. Despite the growing interest in such a unified generative approach for IR systems, the advantages of using a single, multi-task model over multiple specialized models are not well established in the literature. This paper investigates whether and when such a unified approach can outperform task-specific models in the IR tasks of search and recommendation, broadly co-existing in multiple industrial online platforms, such as Spotify, YouTube, and Netflix. Previous work shows that (1) the latent representations of items learned by generative recommenders are biased towards popularity, and (2) content-based and collaborative-filtering-based information can improve an item's representations. Motivated by this, our study is guided by two hypotheses: [H1] the joint training regularizes the estimation of each item's popularity, and [H2] the joint training regularizes the item's latent representations, where search captures content-based aspects of an item and recommendation captures collaborative-filtering aspects. Our extensive experiments with both simulated and real-world data support both [H1] and [H2] as key contributors to the effectiveness improvements observed in the unified search and recommendation generative models over the single-task approaches.

\end{abstract}




\keywords{Generative Retrieval, Generative Recommendation, Joint Search and Recommendation, Multi-task Learning}



\maketitle

\section{Introduction}

\begin{table}[]
\caption{Input and outputs for generative recommendation (\recsmodel{}) and generative retrieval (\searchmodel{}). The id strategy $\phi$ returns the token(s) that identify the item (IDs). The joint model (\jointmodel{}) is trained with both datasets.}
\label{table:joint_model_examples}
\begin{tabular}{@{}lll@{}}
\toprule
 & \multicolumn{2}{c}{Generative recommendation (\recsmodel{}) }\\ \midrule
 & Description & Example \\ \cmidrule(l){2-3} 
Input & Tokens for history items & $[~\phi(item_1), ~\phi(item_2)]$ \\
Output & Tokens for target item & $~\phi(item_3)$ \\ \midrule
 & \multicolumn{2}{c}{Generative retrieval (\searchmodel{})} \\ \midrule
 & Description & Example \\ \cmidrule(l){2-3} 
Input & Tokens for textual query & tokenize(``\textit{brazillian jazz}'')\\
Output & Tokens for relevant item & $~\phi(item_3)$ \\ \bottomrule
\end{tabular}
\end{table}

The recent breakthroughs of Large Language Models (LLMs) have significantly influenced the field of Information Retrieval (IR). In search, where the task is to rank a set of relevant documents for a query, LLMs can be employed to learn better textual representations in both sparse~\cite{formal2021splade,nguyen2023unified,lassance2024splade} and dense retrieval~\cite{izacard2021unsupervised,karpukhin2020dense,behnamghader2024llm2vec}, re-rank items directly~\cite{nogueira2019passage, qin2023large,pradeep2023rankvicuna}, and support evaluation~\cite{faggioli2023perspectives,gilardi2023chatgpt}. Similarly, in recommendation systems, where the objective is to rank a set of relevant items for a user, LLMs can improve user and item textual representations~\cite{wu2021empowering,liu2024once}, generate explanations for recommendations~\cite{silva2024leveraging}, and perform conversational recommendation~\cite{penha2020does,wang2023rethinking}.

However, employing LLMs to retrieve items\footnote{We use the terms \textit{item} and \textit{document} interchangeably in this paper to refer to the units of information that can be recommended and searched for.} for a given query or user is challenging. Architectures based on LLMs such as Cross-encoders~\cite{nogueira2019passage,qin2023large,pradeep2023rankvicuna} lack efficiency for large sets of items
due to their self-attention mechanisms that process the tokens of the query and the tokens of the documents together. Although Bi-encoders~\cite{karpukhin2020dense} and Two-tower models~\cite{yi2019sampling} can employ LLMs for retrieval, they cannot leverage the attention across the query and document tokens~\cite{lin2022pretrained}. Generative retrieval~\cite{de2020autoregressive,tay2022transformer} has been recently introduced to perform retrieval directly within a pre-trained LLM. This method learns to predict document unique identifiers (IDs) for input queries, bypassing the need for separate text-based encodings of queries and documents. Thus, generative retrieval might be effectively applied in a multi-task learning framework~\cite{metzler2021rethinking} to develop a unified model composed of a single LLM, rather than multiple task-specific models. Research indicates that a single model addressing both search and recommendation could enhance the effectiveness of both tasks~\cite{zamani2018joint,zamani2020learning}. However, such a multi-task approach remains under-explored in the generative IR literature. 

This paper seeks to understand the circumstances under which a unified generative model for search and recommendation is beneficial for both tasks. To do so, we study how and when the search task can improve the effectiveness of the recommendation task and vice-versa. Table~\ref{table:joint_model_examples} shows examples of input and output for the recommendation task, in which a generative recommender generates item IDs based on past user interactions, and for the search task, in which a generative retrieval model generates item IDs for a given query. A joint generative model is trained for both tasks.

Previous work shows that (1) the latent representation of items, learned by generative models, are biased towards items popularity\footnote{Popularity of an item is measured by its number of interactions within the recommendation task, or the number of queries leading to it in the search task.}~\cite{liu2023text}, and (2) content-based and collaborative-filtering-based information can improve an item's representations~\cite{yu2012collaborative,thorat2015survey,parthasarathy2023hybrid}. Based on those observations, we formulate two guiding hypotheses for our experiments:  [H1] the joint training regularizes the estimation of each item's popularity, and [H2] the joint training regularizes the item's latent representations. Our main findings based on simulated and real-world data are:

\begin{itemize}
    \item Improvements in the effectiveness of the joint generative model over the task-specific ones are evident in the simulated data for both hypotheses, provided that there is low KL divergence~\cite{kullback1951information} between the popularity distributions of items in search and recommendation for [H1] and that the distribution of item co-occurrences aligns with the other task for [H2], thereby making the regularization effects helpful.

    \item The joint training of generative retrieval models for both recommendation and search proves more effective than the task-specific models for both tasks across three real-world datasets, showing an average increase of 16\% in R@30. Our follow-up analyses suggest that the regularization effect on the item's latent representation (i.e., [H2]) is the primary reason the predictions of the joint generative model differ from those of the task-specific models.
    
\end{itemize}

\section{Related Work}
\paragraph{Generative Retrieval}
Traditional information retrieval methods first encode all documents into a sparse or dense latent representation space and then find the closest document to the query at test time. In contrast, generative retrieval models~\cite{de2020autoregressive,tay2022transformer,zhang2023model,li2023multiview,wang2023novo,yang2023auto,wang2022neural,zeng2023scalable} learn to directly map queries to document IDs, enabling end-to-end pipelines within a single LLM. One of the first retrieval methods to use a Transformer-based language model (LM) is GENRE
~\cite{de2020autoregressive}, which performs entity retrieval to directly output IDs for a query. GENRE predicts the entity name in an auto-regressive fashion, e.g. ``\textit{Leonardo}'' $\rightarrow$ ``\textit{Leonardo da}'' $\rightarrow$``\textit{Leonardo da Vinci}'', using constrained beam search. Generative retrieval has been popularized in IR by DSI~\cite{tay2022transformer}, which uses a LM to predict relevant documents for a given query, where documents are represented by IDs (e.g. \texttt{doc\_01}) rather than its text---leading to fewer tokens representing each document and enables to control tokens that are shared across documents. Considering that we have weights tied to document tokens, generative retrieval requires each document to be explicitly linked to model weights in the LM head.


Tying documents to specific weights in the model requires documents to have been previously observed at training time to be predicted. For this reason, generative retrieval faces challenges of scalability and ingestion of new documents~\cite{kishore2023incdsi,mehta2022dsi++}. Moreover, generative retrieval struggles to outperform and replace existing retrieval approaches when used for larger sets of items~\cite{pradeep2023does}. Recent studies indeed show that additional techniques are required to have competitive generative retrieval approaches~\cite{zeng2023scalable,zeng2024planning}. In this paper, we focus on the effects an additional retrieval recommendation task might have when employing multi-task learning. We expect our results to be agnostic to architectural choices and other explorations in literature aimed at improving the effectiveness and scalability of generative retrieval models.

\paragraph{Generative Recommendation}
While search models retrieve documents based on queries, recommender systems retrieve items based on user's past interactions. These systems may use textual metadata (content-based), historical interactions (collaborative filtering) or both. Traditional models such as the Two-tower model~\cite{yi2019sampling} learn embeddings for both users and items, and use neighbor search to retrieve a candidate set for recommendation, typically followed by a re-ranking step.  In contrast, generative recommender systems can directly retrieve items from a collection for a given user in a single step. Early approaches relied only on user interactions using technologies ranging from LSTMs~\cite{hidasi2015session} to Transformers~\cite{kang2018self} to learn a mapping between users and items. More recently, LLMs have gained popularity for their ability to mix item IDs and natural language~\cite{fan2023recommender,lin2023can}. A notable example is P5~\cite{geng2022recommendation}, which combines text and IDs\footnote{One prompt in P5 could be for example \textit{Given the purchase history list of $user\_15466$: 4110 $\rightarrow$ 4467 $\rightarrow$ 4468 $\rightarrow$ 4472. Find the next item.}} in a pre-trained P5 model to handle multiple recommendation tasks, such as sequential recommendation and rating prediction, within a multi-task setting\footnote{While P5 uses random IDs, more sophisticated content-based ways to create semantic IDs have also been proposed~\citet{rajput2023recommender,hua2023index}, paralleling semantic strategies used to represent document IDs for search. }. While generative search and recommendation have become popular separated research directions~\cite{li2024survey} we study here the effect of learning both in the same model.


\paragraph{Joint Search and Recommendation}
Search and recommendation tasks share similarities, both in terms of the fundamental problem~\cite{belkin1992information}---to identify objects that satisfy users' needs---and in the way models match queries or users (potentially with contextual information) with documents or items~\cite{xu2018deep}. In some domains and platforms such as Spotify, YouTube, and Netflix, items can be accessed through textual search queries or recommendations based on a user's historical interactions and current context. In such cases, a joint search and recommendation model could benefit from the signals coming from both ways to interact with items~\cite{si2023search,wang2012unified}.

\citet{gong2023unified} argued that in industrial applications, the data collected from either search or recommendation scenarios does not fully capture users’ intents, and shows that a multi-task learning approach for seven IR tasks (e.g. Query-Item Relevance Prediction and Content Search Ranking) is more effective in an A/B test for the cold-start problem. Unlike personalized search systems~\cite{dou2007large}, where user previous preferences are used to improve search and models that learn the transitions between search and recommendation surfaces~\cite{shi2024unisar,yao2021user}, we focus here on the problem of learning better item representations with both search and recommendation data. Literature has also hypothesized that sharing training data from both tasks enhances the item representations~\cite{zamani2018joint,zamani2020learning,zhao2022joint}. Although there is evidence for the improved effectiveness of joint learning over individual models, it remains unknown whether this applies to generative models and the underlying reasons each task could benefit from the other. In this paper, we analyze how search and recommendation tasks could benefit from each other in generative retrieval, investigating the learned popularity distribution of items and the shared item representations in generative models.
\section{Research Hypotheses}

\begin{table*}[ht!]

\caption{Motivational examples for our hypothesis. For each row of this table, we have examples of search and recommendation datasets, and the expected effect of training the generative retrieval model on both objectives. Each example motivates a hypothesis we present in the paper. For the second hypothesis, we consider one example for each direction: S $\rightarrow$ R indicates that the target task is recommendation, and thus we expect the search example to improve the joint model effectiveness. Similarly,  R $\rightarrow$ S indicates that the target task is search, and thus we expect the recommendation examples to improve effectiveness compared to a model trained only on search. The images show how the representation of the items could change when going from the task-specific models, $item_i$ (S) or $item_i$ (R) to the joint model $item_i$ (S+R).
}
\label{fig:motivation}
\begin{tabular}{@{}lllp{2.35cm}p{2cm}@{}}
\toprule
\textbf{Hypotheses} & \textbf{Search (S) example} & \textbf{Recommendation (R) example} & \multicolumn{2}{c}{\textbf{Effect R+S joint training}}\\ \midrule
\begin{tabular}{l}
{[H1] Popularity} \\ 
\end{tabular} & 
$Pop^{Train}_{R} = \{5\%, 40\%, 50\%, 5\%\}$ & 
$Pop^{Train}_{S} = \{20\%, 0\%, 50\%, 10\%\}$ & 
\multicolumn{2}{c}{\begin{tabular}{l}$Pop^{Train}_{S+R}  = \{12.5\%, 20\%, 50\%, 7.5\%\}$ \\ gets closer to \\ $Pop^{Test}_{R} = \{10\%, 30\%, 40\%, 10\%\}$ \vspace{0.5em}
\end{tabular}} \\ \hline
\begin{tabular}{l} {[H2] Latent rep.} \\ S $\rightarrow$ R  \end{tabular} & 
\begin{tabular}[c]{@{}l@{}}$q_1$: \textit{``medieval setting''}\\ $q_1$ relevant items: $\{\textcolor{blue}{item_1}, \textcolor{orange}{item_2}\}$\\ \\ $q_2$: \textit{``historical fiction''}\\ $q_2$ relevant items: $\{\textcolor{blue}{item_1}, \textcolor{orange}{item_2}\}$\end{tabular} &
$u_1$: $\{item_3 \rightarrow \textcolor{orange}{item_2} \rightarrow \textcolor{blue}{item_1}\}$&
\includegraphics[height=2.5cm]{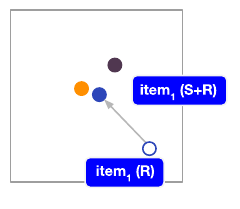}
&
\vspace{-2cm} 
{\footnotesize The embedding of $\textcolor{blue}{item_1}$ using S+R becomes closer to $\textcolor{orange}{item_2}$.} \\
\hline
\begin{tabular}{l}
    {[H2] Latent rep.} \\ { R  $\rightarrow$ S} 
  \end{tabular} &
\begin{tabular}[c]{@{}l@{}}$\textcolor{magenta}{q_3}$: \textit{``ancient history''}\\ $\textcolor{magenta}{q_3}$ relevant items: $\{\textcolor{blue}{item_1}, \textcolor{orange}{item_2}, item_3\}$\end{tabular} &
\begin{tabular}[c]{@{}l@{}}$u_2$: $\{item_3 \rightarrow \textcolor{orange}{item_2}\}$\\ $u_3$: $\{\textcolor{blue}{item_1} \rightarrow \textcolor{orange}{item_2} \}$\end{tabular} &
\includegraphics[height=2.5cm]{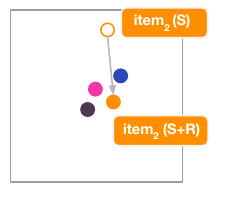}
 & \vspace{-2cm} {\footnotesize The embedding of $\textcolor{orange}{item_2}$ using S+R becomes closer to $\textcolor{blue}{item_1}$ and $item_3$ and consequently to the query $\textcolor{magenta}{q_3}$.} 
\\
\bottomrule
\end{tabular}
\end{table*}

\begin{figure}[] 
\centering
\includegraphics[width=.42\textwidth]{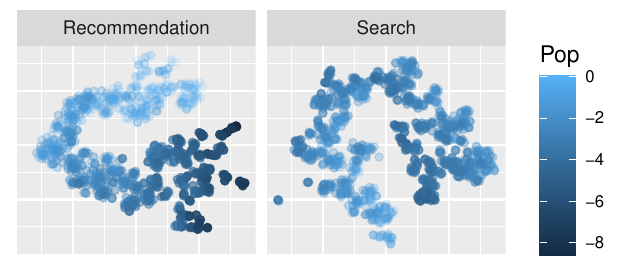}

\caption{t-SNE~\cite{van2008visualizing} projection of the latent representations of a generative recommender (left) and generative search (right) for the ML dataset. The popularity of each item (darker if the item occurs more often in the respective recommendation and search datasets) is encoded in the embeddings of the learned item IDs in both generative retrieval models.
\vspace{-1.5em}
}
\label{fig:popularity_embedding}
\end{figure}

Before defining our first hypothesis, let us discuss our motivation to investigate the popularity of items in generative retrieval. Previous work shows that item IDs in generative recommendation systems tend to exhibit a bias towards more popular items~\cite{liu2023text}. In Figure~\ref{fig:popularity_embedding} we show this bias exists in both search\footnote{Note that many search datasets do not have unequal distributions of queries to documents, such as MSMarco~\cite{nguyen2016ms} and BEIR~\cite{thakur2021beir}.} and recommendation tasks using the latent representations of items (learned by generative retrieval models) in ML, a dataset derived from Movie-Lens 25M \cite{harper2015movielens} (see Section~\ref{sec:experiments}). Considering the importance of item popularity in generative retrieval, our first hypothesis, [H1], is the following:
\begin{hyp}[Popularity]
    A joint model for search and recommendation regularizes the estimation of each item's popularity. 
\end{hyp}

To give a concrete example, let us analyze the first row of Table~\ref{fig:motivation}, which shows our motivational examples to illustrate our main hypotheses. Consider a recommendation dataset comprising four items with training popularity distribution of $Pop^{Train}_{R} = \{20\%, 0\%, 50\%, 10\%\}$ (third column of Table~\ref{fig:motivation}), indicating for example that the first item appears on 20\% of the instances. Let's assume the train distribution differs from the distribution at test time $Pop^{Test}_{R} = \{10\%, 30\%, 40\%, 10\%\}$. This represents a shift from training to test conditions~\cite{quinonero2022dataset}. Suppose that the search data's training set distribution is $Pop^{Train}_{S} = \{5\%, 40\%, 50\%, 5\%\}$, for example because users behave differently across surfaces. A model trained on both datasets (see column R+S from Table~\ref{fig:motivation} for the first row) would retrieve the second item more frequently compared to a model trained solely on $Pop^{Train}_{R}$, leading to better effectiveness on its own shifted test set. For the regularization coming from H1 to increase the effectiveness of the joint model in real-world applications, we hypothesize that the popularity distributions between search and recommendation have to be related but not equal, e.g. low KL Divergence, and the individual datasets have to be insufficient for learning a robust estimation of the test set popularity of each item (e.g. a shift between train and test distributions exists). 

Next, we discuss our second hypothesis motivated by previous work on  content-based and collaborative-based hybrids~\cite{yu2012collaborative,thorat2015survey,parthasarathy2023hybrid} and on joint search and recommendation models~\cite{zamani2018joint,zamani2020learning}:

\begin{hyp}[Latent representation]
    A joint model for search and recommendation regularizes the item's latent representations, meaning that the patterns learned to position items across the latent space in one task can beneficially influence the other.
\end{hyp}

For example, consider the $S \rightarrow R$ task (Search to Recommendation), shown in the second row of Table~\ref{fig:motivation} where recommendation is the target task\footnote{We refer to \textit{target} task as the one we are currently evaluating the joint model on.}. Suppose a scenario in which R aims to predict $\textcolor{blue}{item_1}$ from user $u_1$ who previously interacted with $\{item_3 \rightarrow \textcolor{orange}{item_2}\}$. 
The recommendation model might struggle to make this prediction if the pair $\{item_3, \textcolor{blue}{item_1}\}$ and $\{\textcolor{orange}{item_2}, \textcolor{blue}{item_1}\}$ have rarely (or never) been observed in the training dataset $D^{Train}_{rec}$. However, the search task could fill this data scarcity if $D^{Train}_{search}$ contains semantically similar queries that lead to those three items. Consider for example that, $\textcolor{blue}{item_1}$ is relevant for the queries \{\textit{``medieval setting'', ``historical fiction''}\}, and $\textcolor{orange}{item_2}$ is relevant for the same set of queries. Consequently, the joint model may learn an embedding space where $\{\textcolor{blue}{item_1}, \textcolor{orange}{item_2}\}$ are closer, leading to correctly predicting $\textcolor{blue}{item_1}$ as a likely next item to $\textcolor{orange}{item_2}$. 

Let's now examine an example in the reverse direction $R \rightarrow S$ (third row in Table~\ref{fig:motivation}), where the regularization coming from the recommendation objective might help the search task. Consider a test query $q_3$ \textit{``ancient history''} where the relevant documents are $\{item_3, \textcolor{orange}{item_2}, \textcolor{blue}{item_1}\}$ and that a learned retrieval model correctly retrieves $\{\textcolor{blue}{item_1}, item_3\}$ (they are close to the query in the semantic space). If the pairs $\{\textcolor{blue}{item_1}, \textcolor{orange}{item_2}\}$ and $\{\textcolor{orange}{item_2}, item_3\}$ are prevalent in the $D^{Train}_{rec}$, the joint model will push $item_3$ closer to $\{\textcolor{blue}{item_1}, \textcolor{orange}{item_2}\}$, and consequently to the query, enhancing its retrieval accuracy.

\section{Joint Search and Recommendation Generative Model}
In generative retrieval, a function $\phi$ maps each item in the collection to its respective identifier, which might contain one or more tokens. The vocabulary of the underlying pre-trained LM is composed of the vocabulary tokens that represent textual natural language and of the tokens used to represent the items in the collection. We use atomic IDs\footnote{Future work may explore replacing these atomic IDs with semantic IDs~\cite{tay2022transformer,hua2023index,rajput2023recommender}, based on content or collaborative embeddings, to scale to a larger set of items.} for $\phi$, and thus we have one additional token per item in the vocabulary. Generative models are trained auto-regressively with teacher forcing, employing cross-entropy loss between the predicted ID tokens and the ground truth ID tokens. To perform retrieval with generative retrieval, beam search is performed, returning the top K valid item IDs.

Let $\set{D_S}=\{(Q_i, \{item_1, item_2, ..., item_k\})\}_{i=1}^{N}$ be a search dataset comprised of relevance labels for queries, where $Q$ is the query and $\{item_1, item_k, ..., item_k\}$ are the items that are relevant for this query. To train a generative model on this dataset, each query turns into $k$ input-output pairs of the following format: 
\begin{equation*}
[(Q, ~\phi(item_1)), ..., (Q, ~\phi(item_k))]
\end{equation*}
\noindent
We refer to a generative model trained on  $\set{D_S}$ as \searchmodel{}.  Let $\set{D_R}=\{(U_i = \{item_1, item_2, ..., item_{t-1}\}, item_t)\}_{i=1}^{M}$ be a recommendation dataset comprised of user interactions split into history and target pairs, where the history are the previous interactions of the user sorted by time, and the target item is his last interacted item. To train a generative model on this dataset, each user turns into one pair of the following format: 
\vspace{-1mm}
\begin{equation*}
(concat(\underbrace{[~\phi(item_1), ~\phi(item_2), ..., ~\phi(item_{t-1})]}_{history}), \underbrace{~\phi(item_t))}_{target},
\end{equation*}
where concat$(.)$ is the concatenation of the item IDs with a space token. We refer to a generative model trained on  $\set{D_R}$ as \recsmodel{}. We refer to a single generative retrieval model on both $\set{D_R}$ and $\set{D_S}$ as \jointmodel{}. 
See Table \ref{table:joint_model_examples} for examples.


\section{Experimental setup}
\label{sec:experiments}
In this section, we describe the datasets, the evaluation protocol, and the implementation details we use to analyze the effect of joint training of search and recommendation for generative retrieval.

\paragraph{Simulated datasets}
We create three simulated datasets\footnote{The simulated datasets and their statistics are available at: \url{https://anonymous.4open.science/r/simulated-deatasets-recsys24-DB90/stats.md}} to analyze our hypotheses [H1] and [H2]. For each dataset, we hold the target task data constant and alter the data of the other task based on the specific parameter we want to analyze:
\begin{enumerate}[label=(\roman*)]
    \item \textbf{\sone{}}: This dataset is used to test hypothesis [H1]. We modify the \textit{KL divergence} between the popularity distributions of items to examine how differences in item popularity affect the model's performance. Specifically, we begin by defining a popularity distribution for each item based on Zipf's law~\cite{georgezipf} distribution. Then, we incrementally shuffle the item probability distribution in the search dataset while maintaining a constant distribution in the recommendation dataset. This method introduces increasing divergence between the two popularity distributions. The only learnable signal in both datasets is the popularity bias, allowing us to analyze the impact of one task on the other in a controlled manner, according to how different the popularity distributions are;
    \item \textbf{\stwo{}}: We vary the percentage of queries that lead to the items of the user history which match the queries, i.e. are the same, that lead to the target items (\textit{\% of q. match}) to test [H2] in the $S\rightarrow R$ direction. We simulate a recommendation dataset composed of five clusters, each containing six items that frequently co-occur in user interaction data. User histories are created by randomly sampling an initial item and pairing it with a second item that comes from the same cluster. We do this until the number of desired interactions per user is reached. This setup is designed to encode the co-occurrences of items within specific clusters as the learnable information in the dataset. When generating the search data, we modify how many queries are distributed for items in the same clusters. For example, with \textit{\% of q. match} at 100\% all queries for items within a cluster are the same, whereas, at 50\% overlap, only half of the queries are the same, with the remainder being randomly selected;
    \item \textbf{\sthree{}}: This dataset tests hypothesis [H2] in the $R\rightarrow S$ direction by altering the percentage of item pairs in recommendation data that also appear together in relevant query lists (\textit{\% pairs in qrels}). We simulate a search scenario using a subset of ten randomly selected topics from TREC-DL22~\cite{craswell2023overview}, providing realistic text for the queries. For each topic, we generate five paraphrased queries using OpenAI GPT-4. One of the paraphrases is designated as the test query, while the remaining four serve as training instances.  The recommendation dataset is constructed with varying percentages of appearances of recommendation item pairs in relevant groups of documents (that we refer to as \textit{\% pairs in qrels}). For example, consider a user with the following interactions $\{item_1, item_2, item_3$\} and a search dataset with one query to relevance list $(q, \{item_2, item_3\})$. In this example, the percentage of relevant pairs is 33.33\% as, out of the three pairs coming from all the combinations of items from the user interactions ($\{item_1, item_2\}$, $\{item_1, item_3\}$, and $\{item_2, item_3\}$), only one pair ($\{item_2, item_3\}$) appears in a set of relevant documents for a query.
\end{enumerate} For both recommendation and search tasks, we generate datasets of comparable sizes of training instances. Each model is run five times, and we report the average recall (R@10) values to ensure statistical reliability. We describe in more detail the sampling procedure of each dataset when discussing the results of the simulated datasets.

\paragraph{Real-world datasets} Then, we use three real-world datasets to evaluate the effectiveness of the joint search and recommendation generative model. These datasets are detailed in Table~\ref{table:datasets}.
\begin{enumerate*}[label=(\roman*)]
\item \textbf{ML}: Derived from the MovieLens 25M~\cite{harper2015movielens} dataset, employing user interactions for recommendations, and genres, tags, and genome-tags~\cite{vig2012tag} as search queries. 
\item \textbf{MPD}: Constructed using the Million Playlist Dataset~\cite{zamani2019analysis}\footnote{\url{https://research.atspotify.com/2020/09/the-million-playlist-dataset-remastered/}}, where playlists serve as the recommendation dataset, and unique playlist titles are used as search queries. 

In both ML and MPD datasets, we ensure to have just items that appear in both datasets and that all items in the test set appear in the training set. In addition to these public datasets, we created \item \textbf{Podcasts} dataset to provide an even more realistic assessment of joint search and recommendation effectiveness. This dataset consists of a sample of real log queries and user interactions from Spotify (from 2023-07 to 2023-10), with podcast shows as units of retrieval. This dataset features 84\% of items from the search in the recommendation data, and 30\% of the items in the recommendation data are found in the search data, illustrating how behaviour can vary across different surfaces and samples of datasets. We filtered the search dataset in order to contain mostly broad queries that lead to successful search interactions, as the logs have a high percentage of narrow, known entity queries~\cite{penha2023improving}.

\end{enumerate*}

\begin{table*}[]
\caption{Statistics of the real-world datasets. Avg. rel. per query indicates the average number of relevant documents per query. The Gini index measures how concentrated the items are in the training dataset in terms of queries and users leading to them (higher values mean higher concentration). KSdist is the Kolmogorov Distance between the two distributions of popularities and KLD is the Kullback–Leibler divergence between them.
}
\label{table:datasets}
\begin{tabular}{@{}lcccc
>{\columncolor[HTML]{EFEFEF}}c 
>{\columncolor[HTML]{EFEFEF}}c 
>{\columncolor[HTML]{EFEFEF}}c 
>{\columncolor[HTML]{EFEFEF}}c ccc@{}}
\toprule
\multicolumn{1}{l}{} & \multicolumn{4}{c}{Recommendation} & \multicolumn{4}{c}{\cellcolor[HTML]{EFEFEF}Search} & \multicolumn{3}{c}{Popularity distributions comparison} \\ \midrule

 & \# items 
 & \# users 
 & Density 
 & \begin{tabular}[c]{@{}c@{}}Gini index\end{tabular} 
 
 & \# items 
 & \# queries 
 & \begin{tabular}[c]{@{}c@{}}Avg. rel. \\ per query\end{tabular} 
 & \begin{tabular}[c]{@{}c@{}}Gini index \end{tabular} 
 
 & KSdist(S,R) & KLD(S||R) & KLD(R||S) \\ \midrule
ML  & 20k & 163k & 0.00750 & 0.74 & 20k & 55k & 6.86 & 0.50 &  0.42 & 0.91 & 0.81 \\
MPD  & 21k & 85k & 0.00055 & 0.52 & 21k & 28k & 36.44 & 0.58 & 0.22 & 0.14 & 0.15 \\
Podcasts & 43k & 211k & 0.00030 & 0.73 & 15k & 4k & 10.27 & 0.56 & 0.46 & 1.49 & 1.55 \\ \bottomrule
\end{tabular}
\end{table*}

\paragraph{Evaluation}

In the recommendation task (R), we use the user history to predict the last item (the target) in an ordered sequence of $t$ items. During training, we construct the user's history up to ($t$-3) interactions and simulate additional training instances by sequentially predicting the last item and removing it from the sequence until only two items remain. The test data observes all interactions up to ${
t-1}$ to predict the item at position $t$. The validation test predicts the item at position ${t-1}$ from the previous ones, while the training data does not see the last two interactions for each user.  For the search task (S), we maintain distinct sets of queries for training, validation, and testing. Given our focus on the retrieval task, we rely on recall metrics: R@10 for the simulated datasets\footnote{We used a cutoff of 10 due to the small size of the simulated datasets (<100 entities).} and R@30 for the, larger, real-world datasets. We assess the statistical significance of our results using paired Students' t-tests with a 95\% confidence interval. 

\paragraph{Implementation Details}
We use \textit{Huggingface}'s implementation of T5~\cite{raffel2020exploring} (\textit{google/flan-t5-base}) and train all generative models for 5 epochs with a learning rate of 0.002 and a batch size of 128. We use AdamW~\cite{loshchilov2017decoupled} with a weight decay of 0.01. To increase the number of distinct items retrieved for all generative retrieval models we resort to a diversified beam search approach~\cite{vijayakumar2016diverse}. We use a diversified beam search procedure with a diversity penalty of 0.25. The number of groups equals half the number of desired return items. 

The models used for reference for the search task are BM25~\cite{robertson1994some} (Pyterrier~\cite{macdonald2021pyterrier} implementation with default hyperparameters) and a Bi-encoder~\cite{song2020mpnet} (SentenceTransformers~\cite{reimers-2019-sentence-bert} implementation using pre-trained \textit{sentence-transformers/all-mpnet-base-v2}). The recommendation models used for reference for the recommendation task are SASRec~\cite{kang2018selfattentive} and BERT4Rec~\cite{sun2019bert4rec} using Recbole implementations~\cite{recbole[2.0]}, and DiffRec~\cite{wang2023diffusion} using the author's github code. Although such neural baselines could be referred to as generative recommendation approaches, they do not leverage natural language text tokens, and thus cannot handle the search dataset. Pop$_R$ returns the most frequent items according to the recommendation training data and Pop$_S$ does so according to search.

\section{Results for Simulated Datasets}
This section discusses the results from the simulated datasets. We begin by analyzing the item popularity under [H1] with the first simulated dataset (\sone{}), followed by exploring the regularization effect on item representation for [H2] with \stwo{} and \sthree{}.

\subsection{\sone{}: Item Popularity}
\begin{figure}[t]
\centering
\includegraphics[width=.42\textwidth]{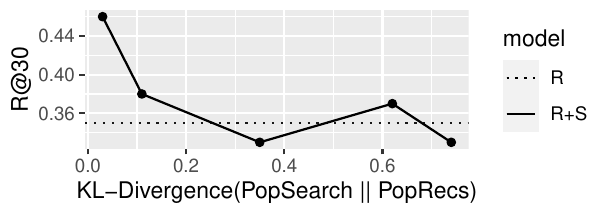}
\caption{Results of \sone{} for [H1]. 
\vspace{-1em}
}
\label{fig:sim_pop}
\end{figure}

We assess [H1] by training generative retrieval models on the datasets from the first simulation, each with varying degrees of KL divergence between item popularity distributions. Figure~\ref{fig:sim_pop} provides evidence supporting [H1]. It suggests that learning for both tasks within a single generative model regularizes the learned popularity distribution of the items. However, we note here that there is no shift between train and test popularity distributions in this simulation. Additionally, the only signal in the simulated data is the popularity of each entity diverging from real world datasets where entities that are similar co-occur more frequently. 

\subsection{\stwo{}: Item Representation (S $\rightarrow$ R)}

\begin{figure}[]
\centering
\includegraphics[width=.42\textwidth]{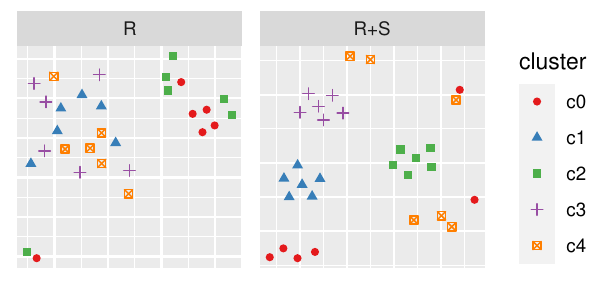}
\caption{t-NSE~\cite{van2008visualizing} reduction of the embeddings of models learned for \stwo{}. 
\vspace{-1em}
}
\label{fig:tsne_sim}
\end{figure}

This section assesses our second hypothesis [H2] by focusing on recommendation as the target task. The results from simulation \stwo{}, presented in Table~\ref{table:res_sim_2_and_3}, show that a mismatch in queries across the clusters negatively affects the regularization, decreasing the effectiveness of \jointmodel{} compared to \recsmodel{}. However, increasing the number of matching queries within clusters enhances the performance relative to the baseline. This finding supports [H2], suggesting that the regularization can be beneficial particularly when there is a high degree of similarity between the search and recommendation data in terms of item co-occurrences (\textit{\% of q. match}). To visually assess the impact of joint training, we compare, in Figure~\ref{fig:tsne_sim}, the item embeddings obtained from a recommendation-only model with the ones obtained by a joint model that incorporates both search (at 100\% of \textit{q. match}) and recommendation data. We observe that the joint model effectively groups items according to their underlying clusters compared to the model trained solely on recommendation.

\begin{table}[]
\caption{Results of \stwo{} and \sthree{} for [H2]. 
$\ddagger$ indicates statistical significance against \recsmodel{} (\stwo{} results) and \searchmodel{} (\sthree{} results) using paired t-tests.}
\label{table:res_sim_2_and_3}
\begin{tabular}{@{}lllllll@{}}
\toprule
Sample & Model & \multicolumn{2}{c}{(\stwo{}) S → R} & Model & \multicolumn{2}{c}{(\sthree{}) R → S} \\ \midrule
& & \begin{tabular}[c]{@{}l@{}}\% of q. \\ match\end{tabular} & R@10 & & \begin{tabular}[c]{@{}l@{}}\% pairs \\ in qrels\end{tabular} & \multicolumn{1}{l}{R@10} \\ \cmidrule(lr){3-4} \cmidrule(lr){6-7} 
\multirow{6}{*}{100\%} & \recsmodel{} & - & 0.907 & \searchmodel{} & - & 0.580 \\ \cmidrule(lr){3-4} \cmidrule(lr){6-7} 
 & \multirow{5}{*}{\jointmodel{}} & 0\% & 0.842  & \multirow{5}{*}{\jointmodel{}} & 0\% & 0.240 \\
 & & 25\% & 0.880 & & 25\% & 0.280 \\
 & & 50\% & 0.938$^\ddagger$ & & 50\% & 0.280 \\
 & & 75\% & 0.944$^\ddagger$ & & 75\% & 0.460 \\
 & & 100\% & \textbf{0.985}$^\ddagger$ & & 100\% & \textbf{0.640}$^\ddagger$ \\ \midrule

\multirow{6}{*}{65\%} & \recsmodel{} & - & 0.717 & \searchmodel{} & - & 0.320 \\  \cmidrule(lr){3-4} \cmidrule(lr){6-7} 
 & \multirow{5}{*}{\jointmodel{}} & 0\% & 0.732 &  \multirow{5}{*}{\jointmodel{}} & 0\% & 0.200 \\
 & & 25\% & 0.752 &  & 25\% & 0.260 \\
 & & 50\% & 0.708 &  & 50\% & 0.220 \\
 & & 75\% & \textbf{0.903}$^\ddagger$  &  & 75\% & \textbf{0.400}$^\ddagger$ \\
 & & 100\% & 0.897$^\ddagger$  &  & 100\% & 0.380$^\ddagger$ \\ \bottomrule
\end{tabular}
\end{table}

\subsection{\sthree{}: Item Representation (R $\rightarrow$ S)}
In this section, we validate our second hypothesis by focusing on search as the target task. Table~\ref{table:res_sim_2_and_3} shows the results for simulation \sthree{}. They indicate that the joint model performs more effectively than the model trained solely on the recommendation data when the \textit{\% pairs in qrels} exceeds a certain threshold. This finding supports [H2], suggesting that the multi-task learning objective can serve as an effective regularizer for item representations. This regularization is particularly beneficial when there is a high degree of similarity between the co-occurrences of items in the search and recommendation data.
\section{Results for Real-World Datasets}
In this section, we present the results from real-world data in a question-and-answer format to provide a structured understanding of the insights derived from the datasets. Table~\ref{table:main_results} shows the main results for the three real-world datasets (Head indicates the effectiveness for the top 1\% most popular items in the train set, where Torso is the remaining set of items.). Note that Table~\ref{table:main_results} also contains baselines (Pop$_R$, Pop$_S$, SASRec~\cite{kang2018selfattentive}, BERT4Rec~\cite{sun2019bert4rec}, and DiffRec~\cite{wang2023diffusion} for the recommendation task and BM25~\cite{robertson1994some} and Bi-encoder~\cite{song2020mpnet} for the search task) as a reference. The aim of this research is not to provide a new generative model that is more effective than the non-generative counterparts but to bridge the gap of understanding when and why jointly training on search and recommendation benefits a \emph{generative} model.

\paragraph{What is the effectiveness of the joint model compared to the task-specific models?}


\renewcommand{\cellalign}{cl}
\begin{table*}[ht]
\caption{R@30 for the joint generative model (\jointmodel{}) compared to the task-specific models (\searchmodel{} and \recsmodel{}), and the respective percentual improvement. Bold indicates the highest score between the joint generative model and the task-specific generative model, and the superscript $\ddagger$ indicates statistical significance against the task-specific model using paired t-tests. Head indicates the effectiveness for the top 1\% most popular items in the train set, where Torso is the remaining set of items.}
\label{table:main_results}
\begin{tabular}{@{}lllllllllll@{}}
\toprule
 & \multicolumn{9}{c}{Recommendation} \\ \midrule
 & \multicolumn{3}{c}{ML} & \multicolumn{3}{c}{MPD} & \multicolumn{3}{c}{Podcasts} \\ \cmidrule(l){2-10} 
 & \multicolumn{1}{l}{All} & \multicolumn{1}{l}{\cellcolor[HTML]{EFEFEF}Head} & \multicolumn{1}{l}{\cellcolor[HTML]{EFEFEF}Torso} & \multicolumn{1}{l}{All} & \multicolumn{1}{l}{\cellcolor[HTML]{EFEFEF}Head} & \multicolumn{1}{l}{\cellcolor[HTML]{EFEFEF}Torso} & \multicolumn{1}{l}{All} & \multicolumn{1}{l}{\cellcolor[HTML]{EFEFEF}Head} & \multicolumn{1}{l}{\cellcolor[HTML]{EFEFEF}Torso} \\ \midrule
\multicolumn{10}{l}{Generative retrieval methods} \\ \midrule

\recsmodel{} & 0.103 & \cellcolor[HTML]{EFEFEF}0.267 & \cellcolor[HTML]{EFEFEF}0.057 & \textbf{0.067} & \cellcolor[HTML]{EFEFEF}0.269 & \cellcolor[HTML]{EFEFEF}\textbf{0.043} & 0.112 & \cellcolor[HTML]{EFEFEF}0.334 & \cellcolor[HTML]{EFEFEF}0.018 \\
\makecell{\jointmodel{} \\ \textit{   improv.}}& \makecell{\textbf{0.119}$^\ddagger$ \\{\color{OliveGreen}(+16\%)}}& \makecell{\cellcolor[HTML]{EFEFEF}\textbf{0.307}$^\ddagger$ \\\cellcolor[HTML]{EFEFEF}{\color{OliveGreen}(+15\%)}}& \makecell{\cellcolor[HTML]{EFEFEF}\textbf{0.066}$^\ddagger$ \\ \cellcolor[HTML]{EFEFEF}{\color{OliveGreen}(+16\%)}} & \makecell{0.055 \\${\color{BrickRed}(-18\%)}$ }& \makecell{\cellcolor[HTML]{EFEFEF}\textbf{0.333}$^\ddagger$ \\\cellcolor[HTML]{EFEFEF}{\color{OliveGreen}(+24\%)} }& \makecell{\cellcolor[HTML]{EFEFEF}0.021 \\\cellcolor[HTML]{EFEFEF}${\color{BrickRed}(-50\%)}$ }& \makecell{\textbf{0.149}$^\ddagger$ \\ {\color{OliveGreen}(+33\%)}} & \makecell{ \cellcolor[HTML]{EFEFEF}\textbf{0.345}$^\ddagger$ \\ \cellcolor[HTML]{EFEFEF}{\color{OliveGreen}(+3\%)} } & \makecell{\cellcolor[HTML]{EFEFEF}\textbf{0.067}$^\ddagger$ \\ \cellcolor[HTML]{EFEFEF}{\color{OliveGreen}(+262\%)} }\\ \midrule

\multicolumn{10}{l}{Popularity and neural-based methods (for reference)} \\ \midrule
\popularityrecs{} & 0.069 & \cellcolor[HTML]{EFEFEF}0.312 & \cellcolor[HTML]{EFEFEF}0.000 & 0.028 & \cellcolor[HTML]{EFEFEF}0.259 & \cellcolor[HTML]{EFEFEF}0.000 & 0.073 & \cellcolor[HTML]{EFEFEF}0.247 & \cellcolor[HTML]{EFEFEF}0.000 \\ 
\popularitysearch{} & 0.062 & \cellcolor[HTML]{EFEFEF}0.264 & \cellcolor[HTML]{EFEFEF}0.005 & 0.026 & \cellcolor[HTML]{EFEFEF}0.241 & \cellcolor[HTML]{EFEFEF}0.000 & 0.029 & \cellcolor[HTML]{EFEFEF}0.091 & \cellcolor[HTML]{EFEFEF}0.003 \\ 
\texttt{SASRec}  & 0.207
&\cellcolor[HTML]{EFEFEF}0.356
&\cellcolor[HTML]{EFEFEF}0.164
& 0.231
&\cellcolor[HTML]{EFEFEF}0.462
&\cellcolor[HTML]{EFEFEF}0.205
&  0.256
&\cellcolor[HTML]{EFEFEF}0.546
&\cellcolor[HTML]{EFEFEF}0.134 \\
\texttt{BERT4Rec}  & 0.247	
&\cellcolor[HTML]{EFEFEF}0.401
&\cellcolor[HTML]{EFEFEF}0.203
&  0.151
&\cellcolor[HTML]{EFEFEF}0.460
&\cellcolor[HTML]{EFEFEF}0.115
&  0.260
&\cellcolor[HTML]{EFEFEF}0.518
&\cellcolor[HTML]{EFEFEF}0.151 \\
\texttt{DiffRec}  & 0.211
&\cellcolor[HTML]{EFEFEF}0.528
&\cellcolor[HTML]{EFEFEF}0.121
& 0.215
&\cellcolor[HTML]{EFEFEF}0.335
&\cellcolor[HTML]{EFEFEF}0.201
& 0.280
&\cellcolor[HTML]{EFEFEF}0.442
&\cellcolor[HTML]{EFEFEF}0.212    \\\midrule \midrule

 & \multicolumn{9}{c}{Search} \\ \midrule
 & \multicolumn{3}{c}{ML} & \multicolumn{3}{c}{MPD} & \multicolumn{3}{c}{Podcasts} \\ \cmidrule(l){2-10} 
 & \multicolumn{1}{l}{All} & \multicolumn{1}{l}{\cellcolor[HTML]{EFEFEF}Head} & \multicolumn{1}{l}{\cellcolor[HTML]{EFEFEF}Torso} & \multicolumn{1}{l}{All} & \multicolumn{1}{l}{\cellcolor[HTML]{EFEFEF}Head} & \multicolumn{1}{l}{\cellcolor[HTML]{EFEFEF}Torso} & \multicolumn{1}{l}{All} & \multicolumn{1}{l}{\cellcolor[HTML]{EFEFEF}Head} & \multicolumn{1}{l}{\cellcolor[HTML]{EFEFEF}Torso} \\ \midrule
 
\multicolumn{10}{l}{Generative retrieval methods} \\ \midrule
\searchmodel{} & 0.020 & \cellcolor[HTML]{EFEFEF}0.053 & \cellcolor[HTML]{EFEFEF}0.000 & 0.028 & \cellcolor[HTML]{EFEFEF}0.040 & \cellcolor[HTML]{EFEFEF}0.002 & 0.080 & \cellcolor[HTML]{EFEFEF}0.148 & \cellcolor[HTML]{EFEFEF}0.004 \\
\makecell{\jointmodel{} \\ \textit{   improv.}} & \makecell{\textbf{0.023}$^\ddagger$ \\ {\color{OliveGreen}(+16\%)} }& \makecell{\cellcolor[HTML]{EFEFEF}\textbf{0.060}$^\ddagger$ \\ \cellcolor[HTML]{EFEFEF}{\color{OliveGreen}(+12\%)} }& \makecell{\cellcolor[HTML]{EFEFEF}\textbf{0.001} \\ \cellcolor[HTML]{EFEFEF} (-) } & \makecell{\textbf{0.033}$^\ddagger$ \\{\color{OliveGreen}(+18\%)}} & \makecell{\cellcolor[HTML]{EFEFEF}\textbf{0.044}$^\ddagger$ \\\cellcolor[HTML]{EFEFEF}{\color{OliveGreen}(+11\%)}} & \makecell{ \cellcolor[HTML]{EFEFEF}\textbf{0.009}$^\ddagger$ \\\cellcolor[HTML]{EFEFEF} {\color{OliveGreen}(+373\%)} }& \makecell{ \textbf{0.106}$^\ddagger$ \\ {\color{OliveGreen}(+33\%)} }&\makecell{ \cellcolor[HTML]{EFEFEF}\textbf{0.171}$^\ddagger$ \\ \cellcolor[HTML]{EFEFEF}{\color{OliveGreen}(+16\%)}} &\makecell{\cellcolor[HTML]{EFEFEF}\textbf{0.034}$^\ddagger$ \\ \cellcolor[HTML]{EFEFEF}{\color{OliveGreen}(+855\%)}}\\  \midrule

 \multicolumn{10}{l}{Sparse and dense methods (for reference)} \\ \midrule
 
\bm{}  & 0.091
&\cellcolor[HTML]{EFEFEF}0.060
&\cellcolor[HTML]{EFEFEF}0.109
&  0.032
&\cellcolor[HTML]{EFEFEF}0.017
&\cellcolor[HTML]{EFEFEF}0.065
&  0.304
&\cellcolor[HTML]{EFEFEF}0.237
&\cellcolor[HTML]{EFEFEF}0.383 \\

\biencoder{} & 0.081
&\cellcolor[HTML]{EFEFEF}0.062
&\cellcolor[HTML]{EFEFEF}0.092
&  0.032
&\cellcolor[HTML]{EFEFEF}0.017
&\cellcolor[HTML]{EFEFEF}0.065
&  0.269
&\cellcolor[HTML]{EFEFEF}0.212
&\cellcolor[HTML]{EFEFEF}0.336 \\ 

\bottomrule
\end{tabular}
\end{table*}


In most scenarios, the joint model is more effective than the task-specific models for search and recommendation tasks. One exception is observed in the MPD dataset, where \jointmodel{} only outperforms the recommendation-specific model (\recsmodel{}) for Head items (R@30 0.333). These results underscore that the search and recommendation tasks mutually benefit each other in generative retrieval, with an average increase of 16\% in R@30.

\paragraph{What are the discrepancies between task-specific and joint model predictions?}

\begin{table}[t]
\caption{Analysis of model predictions for \textbf{[H1]}.
}
\label{table:preds_pop}
\begin{tabular}{@{}lcc@{}}
\toprule
 & Recommendation & Search \\ 
 & $\Delta$ $Pop_{S}$ \recsmodel{} $\rightarrow$ \jointmodel{} & $\Delta$ $Pop_{R}$ \searchmodel{} $\rightarrow$ \jointmodel{} \\\midrule
ML & -2.72\% & +33.23\% \\
MPD & +0.92\% & -7.36 \% \\
Podcasts & -17.46\% & -13.33\% \\ \bottomrule
\end{tabular}
\end{table}

Motivated by [H1], we explore the popularity of items retrieved by the generative models. Specifically, we compute the popularity increase of the items predicted when transitioning from a task-specific model to a joint model. For example, if the target task is recommendation, we measure the increase in popularity according to the search data of the retrieved items by \jointmodel{} compared to the retrieved items by \recsmodel{}. The results, shown in Table~\ref{table:preds_pop}, indicate that joint training of the model does not always lead to an increase in the average popularity of predicted items (according to the new task we add). This could indicate that the effect on the item's latent representation coming from [H2] is entangled and thus interfering with the popularity distribution.


\begin{table}[t]
\caption{Analysis of model predictions for \textbf{[H2]}. 
}

\label{table:preds_regularization}
\begin{tabular}{@{}lcc@{}}
\toprule
 & Recommendation & Search \\
 & \begin{tabular}[c]{@{}c@{}} \textit{$\Delta$ \# matches of history} \\ \textit{$\rightarrow$ target queries}\end{tabular} & \begin{tabular}[c]{@{}c@{}} \textit{$\Delta$ item pairs from R} \\ \textit{in rel pairs for queries}\end{tabular} \\ \midrule
ML & +183.09\% & +136.47\% \\
MPD & +210.09\% & +213.01\% \\
Podcasts & +170.92\% & +314.21\% \\ \bottomrule
\end{tabular}
\end{table}

Building on [H2], we examine the effect of regularization on items' representations within the joint model. Here, we focus on the differences between the predictions of the task-specific and joint models. We measure two metrics: one for the recommendation task and another for the search task. 

For the recommendation task, for each predicted list from \jointmodel{} that is different than \recsmodel{}, given the ground-truth target item $item_t$  and the history $\{item_1,\dots,item_{t-1}\}$, we count all the queries that have both $\{item_i,item_t\}$ ($i<t$) in their relevance set (\textit{$\Delta$ \# matches of history $\rightarrow$ target queries}). For example, if the user history includes $\{item_1,item_2\}$ and the target is $item_3$, and in the search data there are respectively $2$ and $3$ unique queries that have the items $\{item_1,item_3\}$ and $\{item_2,item_3\}$ in their corresponding relevant set, this value would be $5$. 

The second metric,\textit{$\Delta$ item pairs from R in rel pairs for queries}, applicable to the search task (second column of Table~\ref{table:preds_regularization}), calculates the average number of co-occurrences of the pairs of relevant items for a given query in the recommendation data. For example, if the relevant items for a query are $\{item_1,item_3,item_5\}$ and they co-occur in the recommendation data as follows: $\{item_1,item_3\}: 3$, $\{item_1,item_5\}: 4$, $\{item_1,item_5\}: 2$, the average score for the query would be $(3+4+2)/3$. The findings, detailed in Table~\ref{table:preds_regularization}, reveal when the joint model predicts different items than the task-specific models, both metrics we defined increase significantly. This indicates that the regularization of items' latent representation significantly alters the joint model's predictions, providing strong support for our second hypothesis.

\begin{figure}[]
\centering
\includegraphics[width=.42\textwidth]{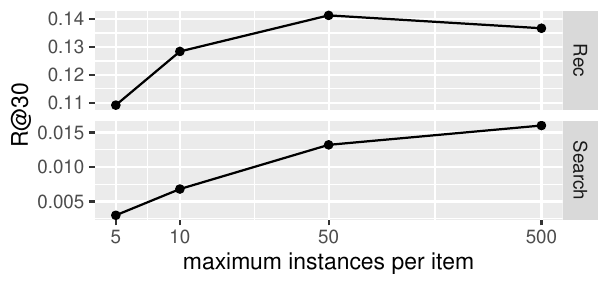}
\caption{Effectiveness of \jointmodel{} for the Podcasts dataset when increasing the number of training instances per item allowed from the added task.
\vspace{-1em}
}
\label{fig:pop_removal}
\end{figure}

\paragraph{What is the effect of removing the popularity information from the added task?}

Although disentangling the effects of both hypotheses from the real-world datasets is challenging, we attempt to remove popularity bias by limiting the number of training instances per item. By restricting the training data to a fixed number of instances per item we effectively remove the popularity bias in the dataset, although we also modify the item representation and co-occurrences for frequently occurring items. The results displayed in Figure~\ref{fig:pop_removal} show a decrease in effectiveness under these conditions, supporting [H1]---that the regularization effect on the learned popularity distribution influences the gains from the joint model.

\paragraph{What is the effect of the redundancy of co-occurring items across the search and recommendation datasets?}

To further understand when the joint model is more effective, we split the test instances based on redundant and non-redundant pairs of items. We define a redundant pair of items if they appear both in the search training data, as relevant for a single query, and in the recommendation training data, as part of the same user history of interactions. A non-redundant pair does not appear in the training set of the target task but appears in the other task training data. While redundant pairs can reinforce useful relationships between pairs, the non-redundant pairs can fill missing patterns between the items that the original task dataset did not have.

The statistics related to the two types of pairs for the Podcasts dataset in Table~\ref{table:redundancy} show that the number of pairs originating from the search dataset, which were not present in the recommendation training data, is relatively low (only 56 non-redundant pairs). This suggests that the observed gains in the Podcasts dataset for the recommendation task are likely not due to learning new relationships between pairs (that were not present in the recommendation data) but rather due to reinforcing the existing ones (31\% improvement for redundant pairs). Conversely, for the search task, we do observe that the gains of the joint model over the task-specific one are higher for non-redundant pairs. This indicates that the regularization effect of [H2] can be beneficial for both redundant and non-redundant pairs of items.

\begin{table}[ht]
\caption{Statistics related to the redundancy of item pairs across tasks for the Podcasts dataset and their effect on finding relevant items. 
}
\label{table:redundancy}
\begin{tabular}{@{}p{4.5cm}cc@{}}
\toprule
 & Recommendation & Search \\ \midrule
Redundant \# pairs & 1933 & 64146 \\ 
Improv. of \jointmodel{} for redundant 
& 31.81\% & 61.39\% \\ \midrule
Non-redundant \# pairs & 56 & 11222 \\ 
Improv. of \jointmodel{} for non-redun.  & 0.00\% & 86.76\% \\ \bottomrule
\end{tabular}
\end{table}
\section{Conclusion}

In this paper, we have explored the impact of multi-task learning in generative retrieval models, specifically focusing on the search and recommendation tasks. We investigated two hypotheses that might explain the effectiveness improvements of multi-task learning for search and recommendation. The first hypothesis is that a joint model can achieve a more robust estimation of each item's popularity, i.e. its frequency of appearance in user interactions or queries. The second hypothesis is that the joint learning objective provides a beneficial regularization effect on the item latent representations.

Using simulated datasets, we provided evidence supporting both hypotheses. We also identify scenarios where improvements might not manifest: e.g. when there is a high divergence between the items' popularity distributions or when the items' co-occurrences do not align across tasks. Additionally, we used three different real-world datasets to show that the multi-task learned generative retrieval model yields effectiveness improvements over the task-specific models, with an average increase of +16\% in R@30. Our analysis detailed the conditions under which the joint model outperforms task-specific models, examining changes in predictions, the impact of eliminating popularity bias in training data and analyzing the effect of item pair redundancy across tasks.

We believe this research marks a significant step towards developing unified LLMs for a broad range of IR tasks, shedding light on the specific contributions of search and recommendation tasks for generative retrieval. For future research, we plan to explore the effect of integrating additional tasks, such as generating explanations, within a unified multi-task learned LLM for IR and to examine ID strategies (how to represent each item based on a set of tokens) in the multi-task learning framework.


\bibliographystyle{ACM-Reference-Format}
\bibliography{sample-base}


\end{document}